\documentclass[aip,apl,amsmath,amssymb,reprint,superscriptaddress]{revtex4-2}
\usepackage{graphicx}
\usepackage{booktabs}
\usepackage{bm}
\usepackage{xcolor}
\usepackage[colorlinks=true,linkcolor=blue,citecolor=blue,urlcolor=blue]{hyperref}
\IfFileExists{ulem.sty}{\usepackage[normalem]{ulem}}{} 
\graphicspath{{"For APL/figs/"}{./figs/}{./}}

\begin{document}

\title[Resonant Raman enhancement in a MoS$_2$ nanotube]{Resonant enhancement of Raman scattering in a MoS$_2$ nanotube}

\author{D.~R.~Kazanov}
\email{kazanovdr@gmail.com}
\affiliation{Ioffe Institute, 194021 St.~Petersburg, Russia}
\author{A.~V.~Poshakinskiy}
\affiliation{Ioffe Institute, 194021 St.~Petersburg, Russia}

\date{\today}

\begin{abstract}
The whispering-gallery modes confined in the wall of a multiwall MoS$_2$ nanotube resonantly enhance the spontaneous Raman scattering by its own optical phonons. The enhancement is doubly resonant --- pump and Stokes waves hit two adjacent modes at one pump energy, and the signal scales as $(Q_{\rm pump}Q_{\rm sc})^2$ --- whenever a phonon energy matches the free spectral range of the mode ladder. Near the A exciton the excitonic dispersion compresses the ladder and sweeps its spacing through all intralayer phonon energies of MoS$_2$, so the match, reached for $A_{1g}$ at $E_{\rm pump}=1551$~meV, requires no geometric tuning. A missed match splits the excitation peak into a pump- and a scattered-resonant doublet whose splitting measures the residual detuning. Selection rules confine the double resonance to the TM$\to$TM channel and exclude $E_{1g}$.
\end{abstract}

\maketitle

Inorganic nanotubes of transition-metal dichalcogenides (TMDs) \cite{Tenne1992,Feldman1995,Remskar2001} inherit\textcolor{red}{,} with a wall of tens of monolayers\textcolor{red}{,} the band structure of the layered 2H bulk \cite{Seifert2000} and thus combine bulk-like excitons and phonons with a rolled geometry, while TMDs themselves have become a high-permittivity platform for nanophotonics \cite{Verre2019,Ermolaev2021}. Raman spectroscopy has accompanied MoS$_2$ from the outset --- the first- and second-order spectra \cite{Verble1970,Wieting1971,Chen1974}, the $E_{2g}^1$--$A_{1g}$ splitting as the layer-number gauge \cite{Lee2010,Li2012,Zhang2015}, the symmetry-dependent resonant scattering near the excitons \cite{Frey1999,Golasa2014,Carvalho2015} --- and for the tubes it probes wall structure, strain, and polytype \cite{Virsek2007,Staiger2012}. All these studies treat the nanotube as a piece of layered material; none considers it as a photonic resonator for its own Raman process.

\begin{figure}[!b]\centering
\includegraphics[width=0.75\columnwidth]{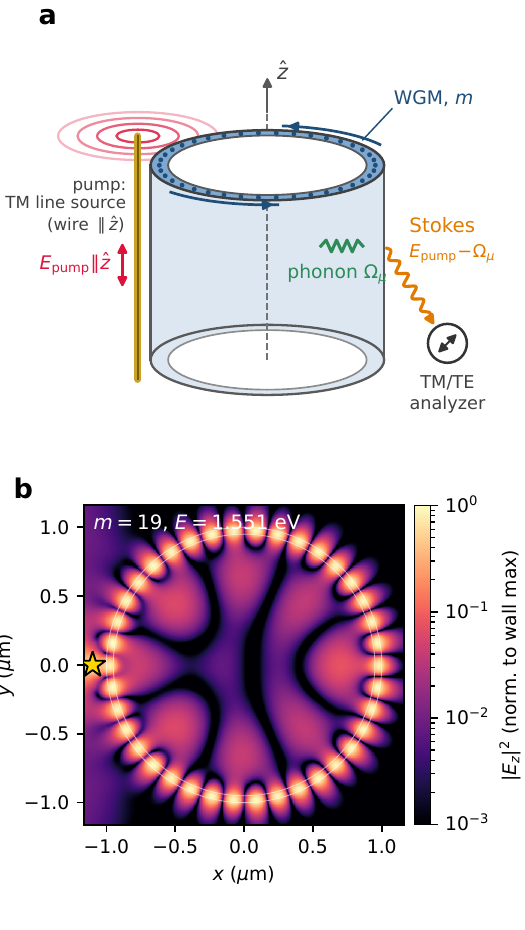}
\caption{(a)~Raman scattering mediated by the wall modes: a TM line source (a wire parallel to the axis, 100~nm from the wall) pumps a whispering-gallery mode through its evanescent field, a phonon $\Omega_\mu$ scatters it into the Stokes wave, and the emission is analyzed in TM or TE polarization. (b)~$|E_z|^2$ of the wall mode $m=19$ at $E=1551$~meV: a ring of $2m=38$ antinodes in the wall; the star marks the line source.}
\label{fig:disp}
\end{figure}

Yet the tube is an excellent resonator: multiwall MoS$_2$ nanotubes support whispering-gallery modes (WGMs) confined within the wall and polarized along the axis \cite{Kazanov2018}, which modulate the excitonic emission \cite{Shubina2019} and whose dispersion we measured along an individual tube \cite{Kazanov2020}. In microresonator photonics, matching the free spectral range (FSR) to the Raman shift --- a double resonance for pump and Stokes --- is the central design requirement of ultralow-threshold Raman lasers \cite{Spillane2002,Rong2005,Latawiec2015}. Here we connect the two lines: we show, analytically and by exact electromagnetic calculation, that the WGMs resonantly enhance the spontaneous Raman scattering of the nanotube's own phonons, maximally when the exciton-compressed FSR matches the phonon energy [Fig.~\ref{fig:disp}(a)].

At normal incidence the problem separates into two polarizations: TM, with $\mathbf E$ parallel to the tube axis and tangential to the layers --- the polarization of the WGMs \cite{Kazanov2018,Shubina2019} --- and TE, with the electric field in the cross section; the subscripts ``pump'' and ``sc'' refer to the pump and the scattered (Stokes) wave.

The central idea is that near the A exciton the dispersion of $\varepsilon_\parallel(E)$ raises the group index $n_g(E)$ of the guided wall wave, so that the intermode spacing set by the $2\pi R$ round trip ($R\simeq1~\mu$m),
\begin{equation}
\Delta(E)\simeq\frac{\hbar c}{R\, n_g(E)}
\label{eq:fsr}
\end{equation}
(the free spectral range, FSR), shrinks and sweeps through the optical-phonon energies of MoS$_2$ ($E_{1g}$ 35.5, $E_{2g}^1$ 47.5, $A_{1g}$ 50.6~meV). When a phonon energy $\Omega_\mu$ equals $\Delta$, the pump and the Stokes wave hit adjacent modes $m$ and $m{-}1$ at one and the same pump energy, and the two resonant enhancements multiply --- a double resonance that arises without geometric fine-tuning.

We model the nanotube as a hollow shell: a MoS$_2$ wall at $R_1<r<R_2$, and vacuum inside ($r<R_1$) and outside ($r>R_2$), with $R_2=1.0~\mu$m and $R_1=0.95~\mu$m.
The MoS$_2$ crystal $c$ axis points radially, so the dielectric tensor is diagonal in the basis of cylindrical coordinates $(r,\phi,z)$:
\begin{equation}
\varepsilon_{rr}=\varepsilon_\perp, \quad
\varepsilon_{\phi\phi}=\varepsilon_{zz}=\varepsilon_\parallel{\color{red},}
\label{eq:eps}
\end{equation}
where the in-plane dielectric permittivity features the A exciton resonance
\begin{equation}
\varepsilon_\parallel = \varepsilon_b+\frac{F_A}{E_A^2-E^2-i\Gamma_A E}{\color{red},}
\end{equation}
with $E_A=1.88$~eV, $\Gamma_A=45$~meV, $F_A=0.75$~eV$^2$, $\varepsilon_b=15.5$, obtained from ellipsometry of bulk MoS$_2$ \cite{Ermolaev2021,Islam2021}, and the out-of-plane permittivity is assumed constant $\varepsilon_\perp=8.41+0.017i$.

\begin{figure}[!t]\centering
\includegraphics[width=0.8\columnwidth]{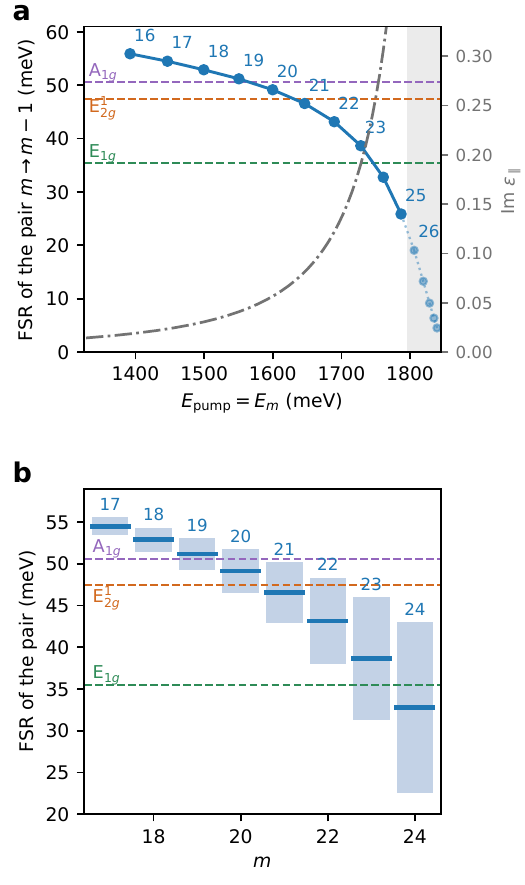}
\caption{Compression and self-tuning of the mode ladder. (a)~FSR of the pair $m\to m{-}1$ at the pump energy $E_{\rm pump}=E_m$ of its upper mode; dashed lines mark the MoS$_2$ phonons. Grey curve (right axis): $\mathrm{Im}\,\varepsilon_\parallel$; in the shaded region $\bar\kappa_m>\Delta$ and the modes merge. (b)~The same FSR values (bars) with their tolerance bands $\delta_{\mathrm{FWHM}}\approx1.3\bar\kappa_m$ (shading): the bands overlap, so each optical phonon lands inside the band of some pair.}
\label{fig:selftune}
\end{figure}


We consider an optical excitation that is homogeneous along the nanotube axis, $k_z=0$. Then, the Maxwell equations separate rigorously into the two scalar polarizations, TE and TM. Next we expand the fields in azimuthal harmonics $\propto e^{im\phi}$ to find the eigenmodes of the system. The incoming and the emitted fields in the Raman problem can be found by solving the boundary-value problem of each harmonic \cite{BohrenHuffman}.


For TM-polarized modes, the nonzero field components are $E_z$, $H_r$, and $H_\phi$. The electric field obeys the wave equation $[\nabla_\perp^2+\varepsilon_\parallel(r)k_0^2]E_z=0$, where $\nabla_\perp^2$ is the Laplacian in the $(r,\phi)$ coordinates and $k_0=\omega/c$. The TM WGMs are governed by $\varepsilon_\parallel$ alone. For them, the 50-nm wall with $\sqrt{\varepsilon_\parallel}\approx4$ acts like a single-mode planar waveguide rolled into a ring. The TM modes of the nanotube with the azimuthal number $m$ are then obtained by enforcing the periodicity condition on the TE modes of the planar waveguide, $n_{\mathrm{eff}}k_0R = m$, where $n_{\mathrm{eff}}$ is the modal index of the guided wave.

The TE-polarized modes of the nanotube have the nonzero field components $E_r$, $E_\phi$, $H_z$ and correspond to the TM modes of the planar waveguide. They have a smaller modal index $n_{\mathrm{eff}}$, which leads to (i) a larger free spectral range and (ii) a slower decay of the evanescent tails and stronger leakage through the centrifugal barrier, resulting in much lower $Q$ factors, as compared to TM modes of the nanotube. In practice, individual TE modes of the nanotube are not resolved at all in the optical spectra, merging into a smooth background.

We therefore focus on the TM modes of the nanotube which have high radiative quality factors. However, due to the absorption, the quality factors are greatly reduced\textcolor{red}{,} $Q_{\mathrm{abs}}/Q_{\mathrm{rad}} \sim 10^{-4}$ for the mode with $m=15$ and $ \sim 10^{-13}$ for $m=24$, so the modes are deeply undercoupled to the free-space modes. Therefore, we choose a near-field source: dipoles on a line parallel to the nanotube\textcolor{red}{,} located 100~nm above its surface, see Fig.~\ref{fig:disp}(a).

The calculated optical spectra reveal the TM-WGM branch $m=15\ldots29$ at $1336$--$1836$~meV ($n_{\mathrm{eff}}=2.2$--$3.1$), see Fig.~\ref{fig:disp}(b) for an exemplary field-intensity distribution. The quality factor $Q_m = 1.2\times10^{3}$ decreases toward the exciton energy, as exciton absorption $\mathrm{Im}\,\varepsilon_\parallel$ grows. In addition, due to the growing exciton contribution to $\mathrm{Re}\,\varepsilon_\parallel$, the FSR compresses from $56.0$ to $6.8$~meV toward the exciton energy, until at $m\simeq25$ the modes merge, see Fig.~\ref{fig:selftune}(a).

The variation of the FSR covers all three intralayer phonon energies, as shown in Fig.~\ref{fig:selftune}(a). That means that under resonant drive of a WGM, the Raman process can resonantly excite a neighboring WGM. Such double-resonant Raman processes are expected to have a greatly increased intensity as compared to non-resonant Raman processes conventionally used in TMDC structures.

However, it is important to note that the set of available FSR values is discrete, so a phonon energy generally cannot match the FSR exactly. Instead, what is required for the double resonance is that the mismatch is lower than the width (FWHM) of the resonances $\kappa$.
The FSR changes by $\Delta{\rm FSR}=-{\rm FSR}^2\, dn_g/(n_g d\omega)$ as the mode number is increased by one. Thus, the tuning of the FSR to a Raman mode frequency is possible if the step $|\Delta{\rm FSR}|$ is smaller than $\omega/Q\approx\kappa$, which leads to the condition $m/Q \gtrsim {\rm FSR}\, |dn_g/(n_g d\omega)|$.
This condition is illustrated in Fig.~\ref{fig:selftune}(b), where the blue lines and intervals indicate the FSR of each mode pair and its tolerance band $\delta_{\mathrm{FWHM}}$, respectively. Clearly, the mode width is larger than the step of the FSR, so all Raman modes, indicated by colored dashed lines, intersect with at least one WGM resonance line.

We now turn to the calculation of the Raman scattering intensity.
A vibrational mode $\mu$ leads to a local modulation of the nanotube wall susceptibility, described by the Raman tensor $\hat R^{(\mu)}$. The Raman tensors for bulk material of the nanotube (with the $D_{6h}$ symmetry) for the three vibrational modes read, in the basis $(\bm e_r, \bm e_\phi, \bm e_z)$,
\begin{equation}\label{eq:RT}
\begin{split}
    R^{(A_{1g})} = \left(\begin{smallmatrix}
         b & 0 & 0 \\
         0 & a & 0\\
         0 & 0 & a
    \end{smallmatrix}\right), \
    R^{(E_{2g}^1)} = \left(\begin{smallmatrix}
         0 & 0 & 0 \\
         0 & -d & 0\\
         0 & 0 & d
    \end{smallmatrix}\right),\
    \left(\begin{smallmatrix}
         0 & 0 & 0 \\
         0 & 0 & d \\
         0 & d & 0
    \end{smallmatrix}\right), \\
    R^{(E_{1g})} = \left(\begin{smallmatrix}
         0 & c & 0 \\
         c & 0 & 0 \\
         0 & 0 & 0
    \end{smallmatrix}\right),\
    \left(\begin{smallmatrix}
         0 & 0 & c \\
         0 & 0 & 0 \\
         c & 0 & 0
    \end{smallmatrix}\right),
\end{split}
\end{equation}
where the two matrices of each $E$ mode correspond to its degenerate components, and $b=1.39\,e^{0.55i\pi}a$, $d=0.66\,a$, $|c|\lesssim0.35\,a$ \cite{Ding2020,Guo2018}; see~Ref.~\onlinecite{Loudon1964}.

Raman vibrations are uncorrelated in different points of the nanotube, so they emit scattered light independently. The total Raman scattering intensity is then given by 
the incoherent sum,~\cite{NovotnyHecht,LeRu2006}
\begin{equation}
I_\mu\propto\!\int\bigl|\mathbf E^{\mathrm{(sc)}}(\omega_{\rm sc})\!\cdot\hat R^{(\mu)}\mathbf E^{\mathrm{(pump)}}(\omega_{\rm pump})\bigr|^2 r\, dr\,d\phi,
\label{eq:If}
\end{equation}
where $\mathbf E^{\mathrm{(pump)}}$ is the incident field at frequency $\omega_{\rm pump}$ generated by the emitter, $\mathbf E^{\mathrm{(sc)}}$ is the detected field at the frequency $\hbar\omega_{\rm sc} = \hbar\omega_{\rm pump} \mp \Omega_{\mu}$, for the (anti-)Stokes scattering by the Raman mode with energy $\Omega_\mu$, and the integral is taken over the nanotube cross-section. We suppose that the detector coincides with the emitter; then $\mathbf E^{\mathrm{(sc)}}(\omega_{\rm sc})$ can be obtained using Lorentz reciprocity: $\mathbf E^{\mathrm{(sc)}}(\omega_{\rm sc})$ $\propto \mathbf E^{\mathrm{(pump)}}(\omega_{\rm sc})$.

\begin{figure*}[t]\centering
\includegraphics[width=0.9\textwidth]{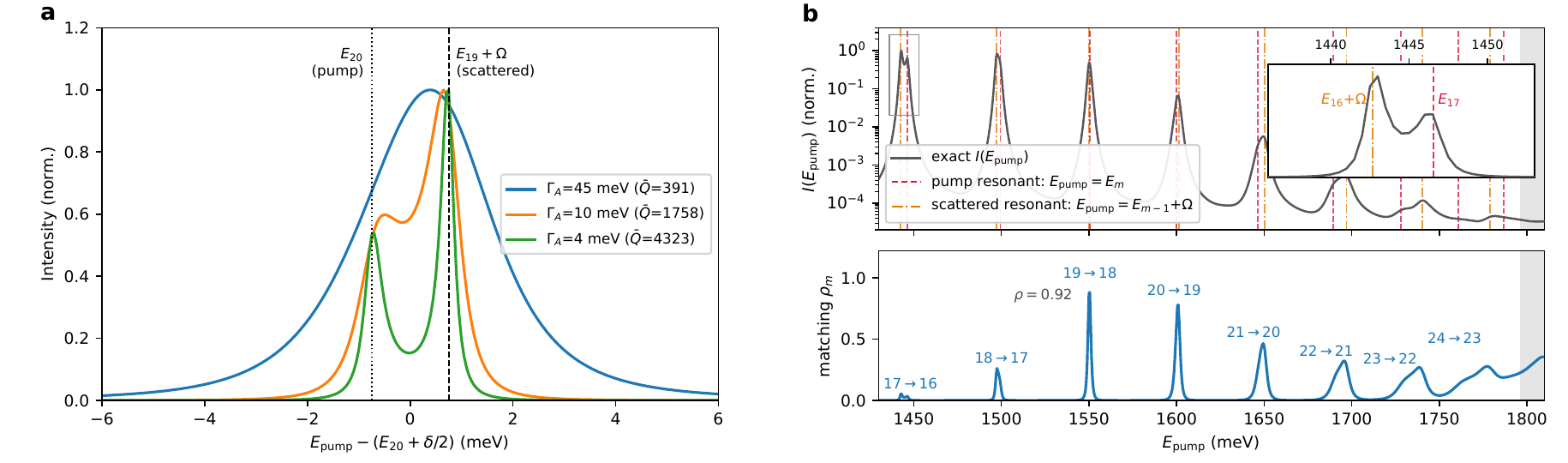}
\caption{(a)~Excitation profile near the pair $m=20\!\to\!19$ ($\delta=+1.45$~meV) for $\Gamma_A=45$, 10, and 4~meV at fixed $E_A$: narrowing $\bar\kappa$ past $|\delta|$ splits the peak into the pump ($E_{20}$) and scattered ($E_{19}+\Omega$) components. (b)~Top: exact excitation spectrum of the $A_{1g}$ line in TM$\to$TM ($\Gamma_A=45$~meV); vertical lines mark the pump resonances $E_m$ (red dashed) and the scattered resonances $E_{m-1}+\Omega$ (orange dash-dotted), which coincide on $19\!\to\!18$ and are 3.9~meV apart on $17\!\to\!16$ (inset). Bottom: factorized profile with unit-height Lorentzians; the peak heights are the matching factors $\rho_m$ of Eq.~\eqref{eq:line}. Grey strips: unresolved modes.}
\label{fig:line}
\end{figure*}

Supposing that both the pump and the scattered field amplitudes in the TM polarization exhibit Lorentz-like resonances, $|\mathbf E^{\mathrm{(pump)}}(\omega_{\rm pump})|^2 \propto 1/[{(\omega_{\rm pump}-\omega_m)^2+\bar\kappa^2/4}]$ and $|\mathbf E^{\mathrm{(sc)}}(\omega_{\rm sc})|^2 \propto 1/[{(\omega_{\rm sc}-\omega_{m\mp1})^2+\bar\kappa^2/4}]$, where $\bar\kappa$ is the mean width of the two modes and we neglect its variation with $m$ for simplicity, we obtain the two possible Raman excitation profiles: (i) If $|\delta| \gtrsim \bar\kappa$, where $\delta= \Omega_\mu-(E_m-E_{m-1})$, there are two resolved lines corresponding to the separate resonances for the incident and the scattered light [see green curve in Fig.~\ref{fig:line}(a)]; (ii) if $|\delta| \lesssim \bar\kappa$ the resonances merge into a double resonance [blue curve in Fig.~\ref{fig:line}(a)]. To differentiate the two cases, we introduce the matching factor, the peak intensity relative to that at exact matching,
\begin{equation}
\rho(\delta) = \frac{\bar\kappa^4}{(\delta^2+\bar\kappa^2})^{2}\,.
\label{eq:line}
\end{equation}
In case (i) $\rho \ll 1$, while in case (ii) $\rho \lesssim 1$ and describes the maximal Raman scattering intensity achieved at $\omega_{\rm pump} = (\omega_m+\omega_{m \mp 1} \pm \Omega_\mu)/2$. The full width at half maximum of $\rho(\delta)$, $\delta_{\mathrm{FWHM}}\approx1.3\bar\kappa$, is the tolerance of the double resonance.

\begin{figure*}[t]\centering
\includegraphics[width=0.85\textwidth]{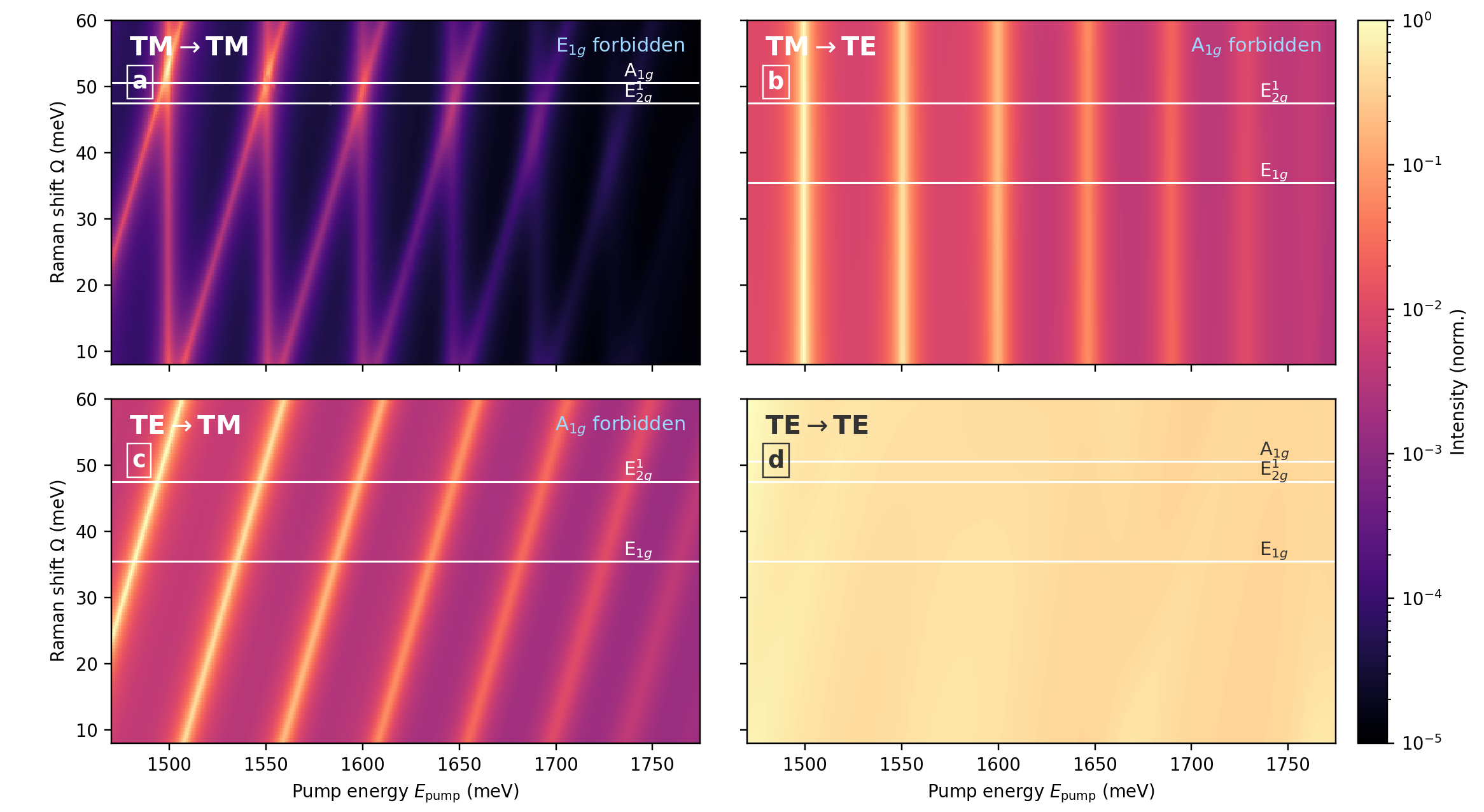}
\caption{Polarization-resolved Raman scattering maps $I(E_{\rm pump},\Omega)$ of the four channels (log scale, each normalized to its maximum; $A_{1g}$ tensor for TM$\to$TM and TE$\to$TE, $E_{1g}$ tensor for the cross maps). (a)~TM$\to$TM: vertical pump-resonant and unit-slope scattered-resonant ridges intersect in the nodes of maximal enhancement; (b)~TM$\to$TE: vertical ridges only; (c)~TE$\to$TM: unit-slope ridges only; (d)~TE$\to$TE: no ridges. White lines mark the phonons allowed in each channel.}
\label{fig:chan}
\end{figure*}

In the real structure, both cases can be realized for different mode numbers, since $\delta$ varies with the mode number: for the $A_{1g}$ mode, $|\delta|$ reaches its minimum of $0.6$~meV for the $19\!\to\!18$ transition, which has $\bar\kappa=3.0$~meV ($\delta_{\mathrm{FWHM}}=3.8$~meV), which gives $\rho = 0.92$, see Fig.~\ref{fig:line}(b); for the $17\!\to\!16$ transition, in contrast, $|\delta|=3.9$~meV exceeds $\bar\kappa=1.7$~meV and the peak is split (inset).

Up to now, we have considered the Raman scattering between two TM modes, which have high quality factors leading to sharp resonances. Now, we compare it with the other polarization channels, see Fig.~\ref{fig:chan}.
The four channels differ in the selection rules. The active Raman modes are marked in the corresponding channel by horizontal lines.

Panel (a) shows the already analyzed scattering map $I(E_{\rm pump},\Omega)$ of the TM$\to$TM channel. It contains vertical ridges (pump resonances, $E_{\rm pump}=E_m$), unit-slope ridges (scattered-wave resonances, $E_{\rm pump}-\Omega=E_{m'}$), and bright nodes at their intersections, where the double resonance condition is fulfilled. The $A_{1g}$ line passes essentially through the node $m=19\!\to\!18$ and the $E_{2g}^1$ line through $m=21\!\to\!20$, while $E_{1g}$ is exactly forbidden in TM$\to$TM.

In the TM$\to$TE channel, only the pump resonances can occur, in the TE$\to$TM, only the scattered-wave resonances can occur, and in the TE$\to$TE no resonances are present, because of the low quality factor of the TE modes.

Table~\ref{tab:peaks} compares all four channels at the matching pump energy of each phonon, normalized to the resonance-free TE$\to$TE channel for the $A_{1g}$ mode, with the Raman tensor elements from Eq\textcolor{red}{.}~\eqref{eq:RT}.
The doubly resonant TM$\to$TM channel of $A_{1g}$ exceeds the reference by $1.5\times10^{4}$ and that of $E_{2g}^1$ by $1.5\times10^{2}$, while the singly resonant cross channels reach only factors of a few. 

\begingroup\squeezetable
\begin{table}
\caption{Enhancement of the four polarization channels at each phonon's matching pump energy (optimum $E_m+\delta/2$), normalized to the resonance-free TE$\to$TE channel of $A_{1g}$, with measured bulk tensor elements \cite{Ding2020,Guo2018}. Zeros are exact symmetry properties; for $E_{1g}$ the nearest pump resonance $E_{23}$ is quoted.}
\label{tab:peaks}
\begin{ruledtabular}
\setlength{\tabcolsep}{3pt}
\begin{tabular}{lccccc}
Mode & $E_{\rm pump}$ & TM$\to$TM & TM$\to$TE & TE$\to$TM & TE$\to$TE\\
\colrule
$A_{1g}$ & 1551 & $\mathbf{1.5\times10^{4}}$ & $0$ & $0$ & $1$\\
$E_{2g}^1$ & 1646 & $\mathbf{1.5\times10^{2}}$ & $2.5$ & $6.1$ & $0.33$\\
$E_{1g}$ & 1728 & $0$ & $\lesssim0.05$ & $\lesssim0.09$ & $\lesssim0.05$\\
\end{tabular}
\end{ruledtabular}
\end{table}
\endgroup

Finally, we discuss the effect of the finite wave vector $k_z$ of the incident and scattered light along the nanotube axis. It shifts the mode energies approximately as $E_m(k_z)\simeq\sqrt{E_m^2+(\hbar ck_z/n_\mathrm{eff}})^2$, see Ref.~\onlinecite{Kazanov2020} for details. However, all the modes with adjacent $m$, involved in the scattering process, shift almost equally, so the FSR drifts by less than $0.5$~meV at the incidence angle $\alpha=30^\circ$, which is much smaller than the typical linewidth $\bar\kappa\approx3$~meV. 
Therefore, the resonance matching conditions are realized simultaneously for many finite $k_z$\textcolor{red}{.}

In summary, a multiwall MoS$_2$ nanotube is a Raman resonator for its own phonons: the exciton-compressed ladder of wall modes provides, without tuning, a pair whose spacing equals the $A_{1g}$ energy, and the doubly resonant TM$\to$TM channel exceeds the resonance-free level by four orders of magnitude. Moreover, the same matching can be used to lower the threshold of stimulated Raman scattering ~\cite{Spillane2002,Rong2005,Latawiec2015}.

\begin{acknowledgments}
This work was partially supported by the State Assignment (Project No.~FFUG-2024-0043). We acknowledge the use of Claude AI in the preparation of the figures and text polishing.
\end{acknowledgments}

\section*{Author Declarations}
\subsection*{Conflict of Interest}
The authors have no conflicts to disclose.
\subsection*{Author Contributions}
\textbf{D.~R.~Kazanov}: Conceptualization; Investigation; Software; Writing -- original draft. \textbf{A.~V.~Poshakinskiy}: Conceptualization; Methodology; Supervision; Writing -- review \& editing.

\section*{Data Availability Statement}
The data that support the findings of this study are available from the corresponding author upon reasonable request.


\end{document}